# Status of the New Research Facility in Andrate (TO) and prospective for research.

D. Tasselli

Department of Astronomy and Astrophysics – TS Corporation S.r.l.
10010 Andrate (TO) – Italy – E-mail: diego.tasselli@tscorporation.org

Keyword: test site – atmospheric event – photometry – new research facility - location: Andrate (TO)

**Abstract**
We report briefly on the status of the new research facility in Andrate (TO) at 839 m above sea level. The structure will be built under the project ROAD073P2T, and will enable the development of major research project. Due to the high altitude, the transparency of the sky, cold, dry and windy at certain times of the year, exceptionally, the location of Andrate (TO) is indeed among the best places in the district of Ivrea (TO), to conduct observation of the sky, in the band U, B, V, R, I, Ir, Rc.

## 1 - Introduction

The new research infrastructure, is located at the City of Andrate - TO in the Salamia Region (coordinates: 45°31'48"07 North, 07°52'28"35 Est.; altitude 839 m). The building will be construct under the project ROAD073P2T part of the first report of new research facilities for the preparation of the Italian part of the European roadmap for research infrastructures. From the observed structure can reach the town of Ivrea (TO) about 10 km from the A4 motorway which move towards the capital Turin and take an airport, or to Milan and reach the Malpensa and Linate airport.

*Fig 1. The location of Andrate (TO), Salamia Region Area, by Google Earth*

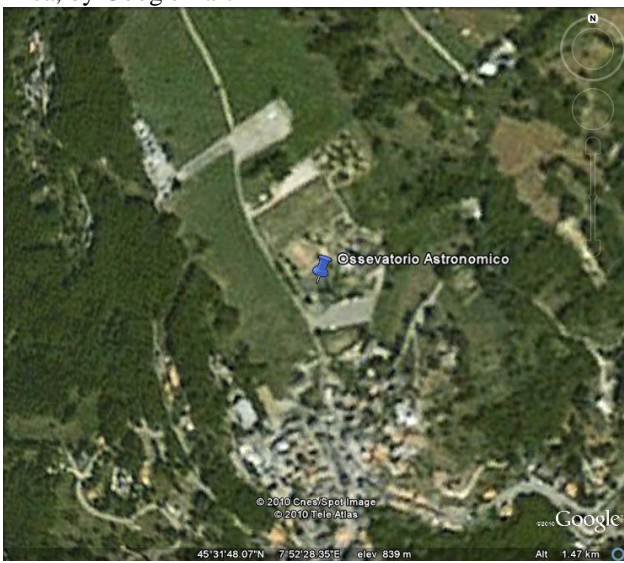

*Table 1. Andrate (TO) coordinates.*

| Geographical coordinates | 45° | 31' | 48".07 | North |
|---|---|---|---|---|
|  | 07° | 52' | 28".35 | East |
| Share | 839 m | | | |

## 2 - Why Andrate

Andrate (TO) is deemed to be the best place in the district of Ivrea (TO) and in the Canavese area, to make observations and researches in the bands U, B, V, R, IR, RC, thanks to its altitude and conditions cold, dry and ventilated with exceptional only in certain times of the year (Figure 2). For these reasons, the atmosphere above the location is very transparent, and the level of noise produced by atmospheric turbulence is very low. In fact, the mists and fogs tend to remain confined, both in summer than in winter, a very low level, which implies a high level of visibility. Therefore, a site with these characteristics, it also offers great benefits in terms of precision astrometry.

*Table 2. Average weather conditions over the past 30 years.*

| Month | T min | T max | Precip. | Humidity | Wind |
|---|---|---|---|---|---|
| gen. | -3 °C | 6 °C | 41 mm | 75% | SSW 4 km/h |
| feb. | -1 °C | 8 °C | 53 mm | 75% | E 4 km/h |
| mar. | 2 °C | 13 °C | 77 mm | 67% | E 4 km/h |
| apr. | 6 °C | 17 °C | 104 mm | 72% | E 4 km/h |
| mag. | 10 °C | 21 °C | 120 mm | 75% | E 4 km/h |
| giu. | 14 °C | 25 °C | 98 mm | 74% | ENE 4 km/h |
| lug. | 16 °C | 28 °C | 67 mm | 72% | ENE 4 km/h |
| ago. | 16 °C | 27 °C | 80 mm | 73% | E 4 km/h |
| set. | 13 °C | 23 °C | 70 mm | 75% | ENE 4 km/h |
| ott. | 7 °C | 17 °C | 89 mm | 79% | E 4 km/h |
| nov. | 2 °C | 11 °C | 76 mm | 80% | E 4 km/h |
| dic. | -2 °C | 7 °C | 42 mm | 80% | SSW 4 km/h |

The main advantages of Andrate (TO) are summarized as follows:

– Exceptional stability (low sky noise);
– A little high-altitude turbulence;

- Sparks bass;
- High precision astrometric and photometric;
- Continuons observation possible;
- Exploitation of natural protection from light pollution produced by the Mombarone's peak of the moraine structure

Preliminary tests on the site were performed with good results (Fig.3-Fig.4). An experimental test was performed to directly compare the short-term atmospheric stability (the so-called sky noise). The data analysis shows a good stability in terms of atmospheric sky noise of different wavelengths.

*Table 3. Seeing data media in the period 2004-2006*

| Year | Values |
|------|--------|
| 2004 | 1,8 arcsec |
| 2005 | 2,0 arcsec |
| 2006 | 1,9 arcsen |

Data on cloud cover to Andrate (TO) has acquired automatically, showed clear skies during most of the winter season, while the average value of maximum wind speed is watched in Table 2.

*Fig 2. Map of the wind speed, in region of Andrate (TO) – Font CESI.*

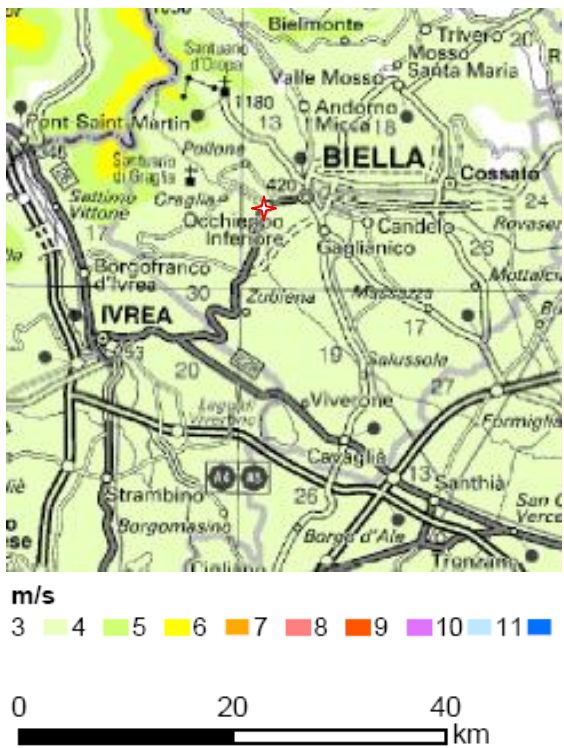

### 3. Observational Station

The research facility will consist of two telescopes placed on either side of the structure with a central area of calculation and handling of electronic instrumentation. The tools have accommodated the following characteristics:

| **Main** |
|---|
| Optical configuration: Ritchey-Chrétien |
| Frame: Equatorial fork |
| Mirrors Main: hyperbolic |
| Precision optics: 1/6 of lambda |
| Useful tool diameter: 600 mm |
| Focal ratio: F8 |
| Weight: 1200 kg |
| **Secondary** |
| Optical configuration: Ritchey-Chrétien |
| Frame: Equatorial fork |
| Mirrors Main: hyperbolic |
| Precision optics: 1/6 of lambda |
| Useful tool diameter: 400 mm |
| Focal ratio: F8 |
| Weight: 900 Kg |
| **CCD Camera and Filter** |
| N. 7 Fly EEV2 back illuminated |
| with Johnson-Cousin filter |
| Diameter standard filters 2" |
| Bands: U, B, V, R, I, Ir, Rc |
| |
| Instruments connected to the main through specific openings arranged under construction and connected with fibber optics. |

The base will be opened later inaugurated 2011, and it will be because they will already operational or technical staff to research. The scientific activities include projects concerning:

- ➢ Solar System: activities and research in asteroids, comets, meteorites and space activities;
- ➢ Stars: activity and research as part of: double stars, cataclysmic variable stars and stars;
- ➢ Cosmology: activities and research as part of galaxies, supernovae, star clusters and Spectroscopy;
- ➢ Meteorology: weather forecasting local and provincial level, modelling climate change to local, national and international;
- ➢ Extra solar planets: activity detection and confirmation of the presence of extra solar planets through the study of their light curves;

The structure with its declared capacity will also be available for current and upcoming projects in astronomy, and expectations can not be anything but very good.

*Fig 3. Seeing graphics from 2004 to 2006 in the new infrastructure area.*

**2004**

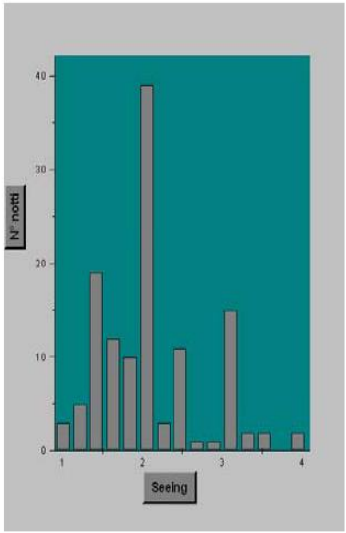

Mean= 1.8 arcsec

**2005**

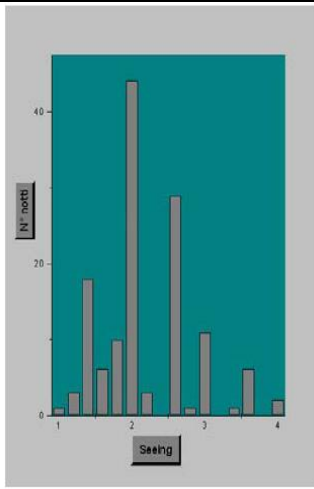

Mean= 2.0 arcsec

**2006**

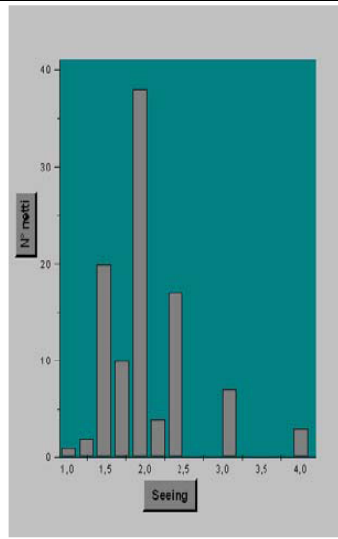

Mean= 1.9 arcsec

*Fig 4. FWHM graphics from 2004 to 2006 in the new infrastructure area.*

**2004**

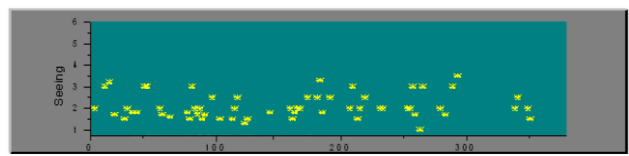

**2005**

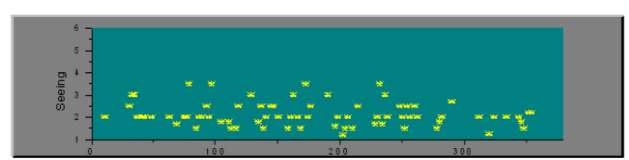

**2006**

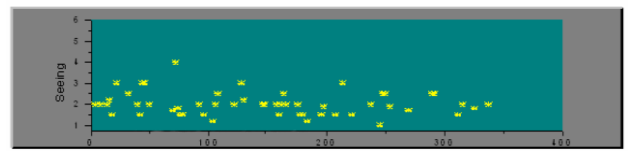